\documentclass{article}
\usepackage{natbib,emulateapj}

\begin{document}

\title{Interpenetrating plasma shells: near-equipartition magnetic field 
generation and non-thermal particle acceleration}

\author{L. O. Silva \altaffilmark{1,6},  R. A. Fonseca \altaffilmark{1}, 
J. W. Tonge \altaffilmark{2}, J. M. Dawson \altaffilmark{2} , W. B. Mori\altaffilmark{2,3}, 
M. V. Medvedev \altaffilmark{4,5}}
\altaffiltext{1}{GoLP/Centro de Fisica de Plasmas, Instituto Superior T\'ecnico, 
 1049-001 Lisboa, Portugal}
\altaffiltext{2}{Dept. of Physics and Astronomy, University of California,
    Los Angeles, CA 90095}
\altaffiltext{3}{Dept. of Electrical Engineering, University of California,
    Los Angeles, CA 90095}
\altaffiltext{4}{Canadian Institute for Theoretical Astrophysics, 
University of Toronto, Toronto, ON, M5S 3H8, Canada}
\altaffiltext{5}{Present address: Department of Physics and Astronomy, University of 
Kansas, Lawrence, KS 66047}
\altaffiltext{6}{email: luis.silva@ist.utl.pt}

\begin{abstract}
We present the first three-dimensional fully kinetic electromagnetic 
relativistic particle-in-cell simulations of the collision 
of two interpenetrating plasma shells. 
The highly accurate plasma-kinetic "particle-in-cell"
(with the total of $10^8$ particles) parallel code OSIRIS has been used. 
Our simulations show:
(i) the generation of long-lived near-equipartition (electro)magnetic fields, 
(ii) non-thermal particle acceleration, and 
(iii) short-scale to long-scale magnetic field evolution, 
in the collision region.  
Our results provide new insights into the  
magnetic field generation and particle acceleration 
in relativistic and sub-relativistic colliding streams of particles, 
which are present in gamma-ray bursters, supernova remnants, relativistic jets, 
pulsar winds, etc..
\end{abstract}


\section{Introduction}

Various violent astrophysical phenomena, such as gamma-ray bursts (GRBs), 
supernova explosions, pulsar wind outflows, knots in relativistic jets, 
are known (or believed) to be rather strong sources of synchrotron radiation 
and cosmic rays, indicating the presence of near-equipartition magnetic fields 
and strong particle acceleration. These conditions are, in general, associated 
with shocks or regions with colliding streams of particles. How the magnetic 
fields are generated, how long do they live and how particles are accelerated 
are still open questions, which can only be definitely addressed via fully 
kinetic three-dimensional numerical simulations. Most astrophysical shocks 
are collisionless because dissipation is dominated by particle scattering 
in turbulent electromagnetic 
fields rather than particle-particle 
collisions \citep{sagdeev}. 
Plasma instabilities driven by 
streaming particles are responsible for the excitation of these turbulent 
electromagnetic fields. The Weibel instability \citep{weibel} has received 
considerable interest as a possible robust mechanism for the production of 
sub-equipartition long-lived magnetic fields and energetic particles in GRBs 
\citep{medvedev,fuller} and pulsar winds \citep{kazimura}. However, 
\citet{gruzinov1} put the simulation results of \citet{kazimura} into question, 
claiming that the produced fields cannot survive far behind the shock.
Previously, Arons and co-workers also pointed out the role of the Weibel 
instability in the downstream region of relativistic magnetosonic shocks 
\citep{arons1,arons2}.

In this Letter, we examine the dynamics of two colliding inter-penetrating 
unmagnetized plasma shells with zero net charge. This is the most simple model 
for the collision region of two plasma shells, as well as a classic scenario 
unstable to electromagnetic and/or electrostatic plasma instabilities.
To probe the full nonlinear dynamics and the saturated state of this system 
it is necessary to employ kinetic numerical simulations. 
We use the fully electromagnetic relativistic particle-in-cell (PIC) 
code OSIRIS \citep{hemker,fonseca} to perform the first three-dimensional 
kinetic simulations of the collision of two plasma shells, and to observe the 
three-dimensional features of the electromagnetic filamentation instability, 
or Weibel instability. 
We compare the collision of 
weakly-relativistic and ultra-relativistic plasma shells. We also point out 
the most distinct features of the three-dimensional simulations, not present 
in lower dimensional simulations. Finally, we discuss the key points raised 
by this work and the directions for future work.

\section{Simulation model} 
In this paper we illustrate the main features of the collision of two plasma 
shells, using the PIC code OSIRIS \citep{hemker,fonseca}. A general description 
of PIC codes is presented by \citet{dawson} and \citet{birdsall}. This scenario 
is pervasive in astrophysical scenarios, and it is of particular relevance in 
the early formation stages of collisionless shocks. Our simulations are directly 
relevant to internal shocks in gamma ray bursts, in connection with collisions of 
electron-positron fireball shells.   

The simulations were performed on a $256 \times 256 \times 100$ grid 
(the axes are labeled as $x1,\ x2,\ x3$) with a total of 105 million particles 
for 2900 time steps, with periodic boundary conditions. A parameter scan 
was done in order to guarantee the grid resolution 
and the size of the simulation box do not affect the simulation results. 
Temporal and spatial scales in the simulations are normalized to the inverse 
electron plasma frequency $\omega_{pe}^{-1}=(4\pi e^2n_{e}/m_e)^{1/2}$,
and the collisionless skin depth $\lambda_{e}=c/\omega_{pe}$, 
where $n_{e}$ is the electron-positron shell density; charges 
and masses are measured in the units of the electron charge $e$ 
and the electron mass $m_e$. In these normalized units, the box size is 
$V=25.6 \times 25.6 \times 10.0 \left( c/\omega_{pe}\right)^3$, and the 
simulations ran for 150.0 $\omega_{pe}^{-1}$.
In all runs, energy is conserved down to 0.025 \%.
In our simulations, a charge-neutral plasma shell, consisting of electrons 
and positrons and moving with a bulk momentum $u_{3} = \gamma_0 v_{3}/c$ along 
the $x3$ direction (vertical direction), penetrates into a similar plasma shell, 
moving in the opposite direction with the same momentum $u_{3}$ (here 
$\gamma_0$ is the initial Lorentz factor of the particles and $v_{3}$ is the 
bulk velocity of a shell along $x3$). Thus, at $t=0$ there are two groups of 
particles moving in opposite directions in the center of mass frame and
occupying the entire simulation volume. 
The particles in both groups (shells) have a thermal spread with the rms 
momentum $u_{th}=\gamma_0 v_{th}\simeq 0.1 $. The system has no net charge and no
net current, and initially the electric and magnetic fields are set to zero. 
We performed both sub-relativistic 
($u_{3}=0.6$, $\gamma_0 \approx 1.17$, $v_{th}=0.085c$) and ultra-relativistic
($u_{3}=10.0$, $\gamma_0 \approx 10.05$, $v_{th}=0.01c$) simulations.
By using the mass ratio of 1 (electron-positron plasma), we can follow more 
clearly the crucial features of the instability process. Note that in the 
dimensionless units used, the results of the simulations are equally 
applicable to the collision of proton -- anti-proton shells as well.
We have also performed collisions of electron-proton plasma shells 
and we have observed a two-stage instability process, with each stage
evolving on the electron and proton collisionless time scales, 
following the pattern observed for the electron-positron shells 
(see \citet{medvedev}, and \citet{tonge}).  

\section{Numerical results and physical picture}
The fundamental issue to be addressed here is the level of the electromagnetic 
field generated via plasma instabilities during the collision of two plasma 
shells, as well as the saturated state of the particles and fields. As we show, the growth 
rate is so short that the saturated state is all that matters 
for astrophysical conditions. 
In Figure \ref{fig1}, we present the temporal evolution of the total energy in 
the produced magnetic and electric fields normalized to the initial 
total kinetic energy in the system ($\epsilon_p$), 
$\epsilon_B=\int B^2 dV/8\pi\epsilon_p$ and
$\epsilon_E=\int E^2 dV/8\pi\epsilon_p$, for the sub-relativistic ($u_{3} = 0.6$) 
and ultra-relativistic ($u_{3} = 10$) scenarios. In our normalized units, 
the initial total kinetic energy of particles in the simulations is 
$\epsilon_p=4\times(\gamma_0-1)\times10.0\times(25.6)^2 $. During the linear
stage of the instability we observed the rapid generation of a strong magnetic 
field, which predominantly lies in the $x1x2$--plane, i.e.,  
perpendicular to the direction of motion of the plasma shells. The magnetic 
field energy density reaches $\sim$~5\% of $\epsilon_p$ for $u_{3} = 0.6$ and  
$\sim$~20\% for $u_{3} = 10.0$. In all cases, the produced electric field is  
significantly weaker than the magnetic field. After the linear stage, the 
instability saturates and the energy in the magnetic field decays rapidly 
(on the collisionless time scale $\omega_{pe}^{-1}$), until it reaches a 
quasi-steady level with no or very slow decay on a time scale much longer 
than $\omega_{pe}^{-1}$. These features are present in both the weakly 
relativistic and ultra-relativistic conditions. Such long-lived magnetic 
fields, associated with the Weibel instability, have also been observed in 
laboratory experiments \citep{clayton}. The linear growth rate agrees well
with the theoretical estimates for the full electromagnetic instability 
(maximum growth rate, $\Gamma$, given by 
$(\Gamma/\omega_{pe})^2 \sim 2 \beta^2_0/\gamma_0(1+\beta_\mathrm{th})^2 $, 
for $\gamma_0 \gg 1$, and $\beta_\mathrm{th} \ll 1$) \citep{silva1}.
Figure \ref{fig1} also illustrates the main features of the Weibel instability, 
in particular, the linear growth rate scaling with $\gamma_0^{-1/2}$ 
and the scaling of the saturation level of the magnetic field with 
$v_{th}^{-1}$. The saturation level of the magnetic field, $B_\mathrm{sat}$, i.e., 
when B-field growth stalls, can be determined by 
combining the analytical results for $\Gamma$, and an estimate 
for the bounce frequency of the particles, $\omega_\mathrm{bounce}$, in the magnetic wells 
generated by the Weibel instability. 
Saturation occurs when $\Gamma \sim \omega_\mathrm{bounce}$, 
yielding 
$B_\mathrm{sat} \sim \sqrt{8 \pi} m_p^{1/2} n_e^{1/2} 
\beta_0 c \gamma_0^{1/2}/(1+\beta_\mathrm{th})$, 
where $m_p(q_p)$ is the mass(absolute charge) of the particles in the cloud. 
Electron-proton clouds will drive higher levels of saturated B-field, 
by a factor of $(m_p/m_e)^{1/2}$ \citep{tonge}, 
but on a longer time scale since $\Gamma \propto m_p^{-1/2}$. 
However, $\epsilon_B$ is independent of $n_e$, $\gamma_0$ and $m_p(q_p)$ 
(see also \cite{medvedev} for a discussion). 
Thermal effects are known to play an important role in the evolution of the Weibel instability 
\citep{silva1}. Higher plasma temperatures will lead to a decrease of 
$\Gamma$ and $B_\mathrm{sat}$, and longer transverse wavelength filaments, 
 within the orders of magnitude observed in our simulations (Silva et al. in preparation).
The final steady-state level of the magnetic 
field is still quite high, up to 0.25\% of the initial total kinetic energy 
in the sub-relativistic case and an order of magnitude higher in the 
ultra-relativistic conditions. This residual magnetic field is mostly
perpendicular, i.e., the $x1$ and $x2$ components are dominant.

Physically, the inhomogeneities in the current will generate inhomogeneities in the 
magnetic field, which in turn will enhance the current inhomogeneities (thus closing 
the instability feedback loop), and generating a large number of current filaments with 
oppositely directed currents. The magnetic field associated with these currents has
also a filamentary structure, as illustrated in Figure \ref{fig2}.
In the early stages of the instability, one observes randomly distributed 
current and magnetic filaments. Figure \ref{fig2}a shows small-scale tilted 
iso-levels of the magnetic field energy density. When the instability enters 
the saturation phase ($t \approx 15-30 \ \omega_{pe}^{-1}$), current filaments
begin to interact with each other, forcing like currents to approach each
other and merge. During this phase, initially randomly oriented filaments cross each 
other to form a more organized, large-scale quasi-regular pattern. 
The strong decrease in the magnetic field 
energy is associated with a topological change in the structure of currents and 
magnetic fields, cf. Figure \ref{fig2}a, \ref{fig2}b, and \ref{fig2}c. 
After saturation ($t \gtrsim 30\ \omega_{pe}^{-1}$), 
the filament coalescence continues, as indicated by the increase 
of the correlation scale, $\sim k^{-1}$, of the $B$-field in
Figure \ref{fig3}. However, the spatial distribution of currents is now quite 
regular, so that filaments with opposite polarity no longer cross each other 
but simply interchange, staying always far away. The total magnetic field 
energy does not change any more. Note that the residual 
magnetic field is highly inhomogeneous, seen as a collection of magnetic 
field domains or "bubbles".  The amplitude of the field in the bubbles is
close to equipartition. Therefore, the overall decrease of the $B$-field
energy is mostly associated with decreasing {\it filling factor} of the
field. Note also that the magnetic domains separate current filaments
of opposite polarity. The discussed temporal evolution can be easily 
followed in the magnetic field energy spectral density, Figure \ref{fig3}. 
The small-scale magnetic field is generated in the linear stage of the 
instability, attains its maximum value, and after saturation it evolves to 
larger length scales (i.e., smaller $|k|$) with the characteristic 
$e$-folding time $\sim\omega_B=eB/m_ec$ (Medvedev, in preparation);
note that $\omega_B/\omega_{pe}=\sqrt{2\epsilon_B}$ and this is valid for 
relativistic particles as well.  

The topological evolution of the magnetic field is accompanied by strong 
heating and non-thermal particle acceleration, as illustrated in Figure 
\ref{fig4}. Generation of non-thermal fast particles is more pronounced in 
the highest $B$-field scenarios. We have observed thermal (rms) momenta 
$u_{th}$ as high as the initial bulk momentum $u_{3}$. The magnetic field 
energy grows in the early stage by slowing down the plasma shells. 
Saturation is achieved by the combination of transverse energy spread 
and near-equipartition magnetic field generation. After saturation, 
the energy stored in the magnetic field is transferred back to the plasma 
particles, leading to strong heating and the generation of high energy 
tails in the distribution function, with energies up to $4 (\gamma_0-1)$ 
for both sub-relativistic and ultra-relativistic plasma shells. 
The presence of such high-energy particles is fundamental to provide 
the mildly relativistic particles to be injected in other accelerating structures, 
such as collisionless shocks. Comparison of Figures \ref{fig3} 
and \ref{fig4} also shows that non-thermal particle acceleration and strong 
plasma heating are correlated with the topological evolution of the magnetic 
field when it evolves from small-scale structures to large-scale structures. 

To elucidate the difference between the evolution in 
three dimensions vs. two dimensions, we performed several 2D runs using the 
same code: the runs in a plane parallel to the 
streaming velocity lead to an under-estimation of the magnetic field energy, 
while 2D runs in the plane perpendicular to the streaming direction of motion
show suppression of the excitation of relativistic plasma waves along this 
direction and, hence, overestimate the magnetic field energy.

A critical question for the validity of the GRB synchrotron shock model
(see, e.g., a review by \citealp{Piran99}) is whether the magnetic field 
produced in the collision region survives for a sufficiently long time (e.g., 
comparable to the synchrotron cooling time), typically 
$\sim10^5\omega_{pe}^{-1}$ for prompt GRB emission and early afterglow 
and up to $\sim10^{10}\omega_{pe}^{-1}$ for late (radio) afterglow, 
as measured in the collision frame. 
Our simulations show very weak or no
field dissipation on time-scales $\sim 150\ \omega_{pe}^{-1}$ 
in the collision region. 
In principle, on much longer time-scales, the $B$-field may be destroyed
via reconnection. The back-of-the-envelope estimate of the reconnection
time yields: $t_{rec}\sim(v_{rec}k)^{-1}\sim(0.1v_Ak)^{-1}\sim
(0.1\epsilon_B^{1/2}k\lambda_e)^{-1}\sim100(k\lambda_e)^{-1}$,
where $v_{rec}\sim 0.1v_A$ is the reconnection velocity \citep{reconn} and
$v_A=B/(4\pi m_e n_e)^{1/2}$ is the electron Alfv\'en speed. Since $k$ decreases
with a shorter $e$-folding time $\sim1/\sqrt{\epsilon_B} \sim 10$, 
reconnection slows down at large times and the field is able to survive
for quite a while. In addition, a fundamental three-dimensional feature 
is the small tilting of current filaments.  
This is intrinsically three-dimensional and it is a manifestation of 
the coupling between the electromagnetic modes ({\bf k} in the $x1x2$--plane) 
and the electrostatic modes ({\bf k} parallel to the streaming velocity of the clouds), 
making magnetic field reconnection much more complicated 
(Silva et al. in preparation). 

\section{Conclusions}
We presented the first self-consistent three-dimensional simulations of the 
fields present in collisions of plasma shells  where the electromagnetic 
two-stream (or Weibel) instability develops. Our results demonstrate that 
this instability in three-dimensions is able to generate sub-equipartition 
quasi-static long-lived magnetic fields on the collisionless temporal and 
spatial scales in the collision region, giving credence 
to the predictions of \citet{medvedev} and 
\citet{fuller}. After the linear stage of the instability, we first observe the decay of the 
magnetic field energy, as also observed by \citet{gruzinov2}, followed by the 
evolution to a residual saturated magnetic field energy density. 
These fields maintain a strong saturated level on 
time-scales much longer than collisionless, at least for the duration of 
the simulations. The next step is to increase the simulation region spatial dimension 
in order to determine the spatial spread of the generated field. 
The obtained values of the equipartition parameter 
($\epsilon_B\sim 3\times10^{-2} - 3\times10^{-3}$  for ultra- and 
sub-relativistic shocks) agree well with the values of $\epsilon_B$ inferred 
from GRB afterglows \citep{PK02}. Furthermore, we show that strong transverse 
temperature increase and non-thermal particle acceleration 
occur when the instability saturates. The generated magnetic 
field evolves from small wavelengths to long wavelengths. Such a behavior 
may explain the observed evolution of the soft spectral index $\alpha$
in time-resolved GRB spectra (\citealp{GCG02}; Medvedev (in preparation))
as a jitter-to-synchrotron spectral transition \citep{Medvedev00}.
Our results indicate that the fields necessary in the early formation stages  
of a shock front  can be easily created 
via plasma instabilities of streaming plasmas. These simulations open 
the way to the full three-dimensional PIC modeling of relativistic 
collisionless shocks, which necessarily involve a different simulation geometry, 
and will be presented elsewhere.

\acknowledgments
This work was partially supported by FCT (Portugal) 
through grants PESO/PRO/40144/2000 and 
POCTI/33605/FIS/2000, and grants DOE DE-FGO3-98DP0021 and NSF PHY-0078508.

\begin{figure}
\caption{Temporal evolution of the total energy in 
$B_{\perp}=(B_{1}^2+B_{2}^2)^{1/2}$ and in $E_3$, normalized to $\epsilon_p$. 
Energy on other components, i.e., in $B_3$ and 
$E_{\perp}=(E_{1}^2+E_{2}^2)^{1/2}$ is much smaller.
\label{fig1}}
\end{figure}

\begin{figure}
\caption{Current filamentation: magnetic field energy density in weakly relativistic scenario 
($\gamma_0=1.17$): (a) slightly before saturation, $t=13.52\ \omega_{pe}^{-1}$, 
(b) after saturation $t=41.60 \ \omega_{pe}^{-1}$, for an iso-surface at 20~\% 
of the maximum energy density; in (c) mass density after saturation $t=41.60 \ \omega_{pe}^{-1}$
for an iso-surface at 75 ~\% of the maximum electron density for species with initial positive current along $x3$ ($j_3 > 0$) (red iso-surface) and $j_3 < 0$ (blue iso-surface). Color bar in (a) and (b) represents magnetic field energy density (in simulation units) - plotted only values above the isosurface level (@ 20\% of maximum).   
\label{fig2}}
\end{figure}

\begin{figure}
\caption{Temporal evolution of the magnetic field energy spectral density 
distribution, normalized to the peak  spectral density ($\gamma_0=1.17$). 
\label{fig3}}
\end{figure}

\begin{figure}
\caption{Temporal evolution of the energy distribution of plasma particles 
(color scale $\equiv$ absolute value of charge density in the simulation units).
\label{fig4}}
\end{figure}

\begin{thebibliography}{}
\bibitem[Birdsall \& Langdon(1982)]{birdsall} Birdsall, C. K., \& Langdon, A. B. 1985, Plasma Physics via Computer Simulation (New York: McGraw-Hill) 
\bibitem[Biskamp \& Schwarz(2001)]{reconn} Biskamp, D., \& Schwarz, E. 2001, Phys. Plasmas, 8, 4729
\bibitem[Dawson(1980)]{dawson} Dawson, J. M. 1983, Rev. Mod. Phys., 55, 403
\bibitem[Fonseca et al.(2002)]{fonseca} Fonseca, R. A., et al. 2002, Lecture Notes in Computer Science 2329, III-342 (Heidelberg: Springer-Verlag)
\bibitem[Gallant et al.(1989)]{arons2} Gallant, Y., et al. 1992, \apj, 391, 73
\bibitem[Ghirlanda et al.(2002)]{GCG02} Ghirlanda, G., Celotti, A., \& Ghisellini, G. 2002, astro-ph/0206377
\bibitem[Gruzinov(2001a)]{gruzinov1} Gruzinov, A. 2001, astro-ph/0111321
\bibitem[Gruzinov(2001b)]{gruzinov2} Gruzinov, A. 2001, \apjl, 563, L15
\bibitem[Hemker(2000)]{hemker} Hemker, R. G. 2000, Ph.D. Thesis, UCLA
\bibitem[Kazimura et al.(1998)]{kazimura} Kazimura, Y., et al. 1998, \apjl, 498, L183
\bibitem[Lal et al.(1997)]{clayton} Lal, A. K., et al. 1997, Phys. Plasmas, 4, 1434
\bibitem[Langdon et al.(1988)]{arons1} Langdon, A. B., et al. \prl, 61, 779
\bibitem[Medvedev(2000)]{Medvedev00} Medvedev, M. V. 2000, \apj, 540, 704
\bibitem[Medvedev \& Loeb(1999)]{medvedev} Medvedev, M. V., \& Loeb, A. 1999, \apj, 526, 697 
\bibitem[Panaitescu \& Kumar(2002)]{PK02} Panaitescu, A., \& Kumar, P. 2002, \apj, 571, 779
\bibitem[Piran(1999)]{Piran99} Piran, T. 1999, Phys. Rep., 314, 575
\bibitem[Pruet et al.(2001)]{fuller} Pruet, J., Abazajian, K., \& Fuller, G. 2001, \prd, 64, 063002  
\bibitem[Sagdeev(1966)]{sagdeev} Sagdeev, R. Z. 1966, Rev. Plasma Phys., 4, 23
\bibitem[Silva et al.(2002)]{silva1} Silva, L. O., et al. 2002, Phys. Plasmas, 9, 2458
\bibitem[Tonge(2002)]{tonge} Tonge, J. W. 2002, Ph.D. Thesis, UCLA 
\bibitem[Weibel(1959)]{weibel} Weibel, E. S. 1959, \prl, 2, 83 
\end{thebibliography}
\end{document}